\documentclass[11pt]{article}
\usepackage{moriond,epsfig,amsmath,amssymb}

\def\be{\begin{equation}}
\def\ee{\end{equation}}
\def\bea{\begin{eqnarray}}
\def\eea{\end{eqnarray}}

\begin{document}
\vspace*{4cm}
\title{OVERVIEW OF CONSTRAINTS ON NEW PHYSICS IN RARE $B$ DECAYS}

\author{DAVID M. STRAUB}

\address{Scuola Normale Superiore and INFN, Piazza dei Cavalieri 7,\\
56126 Pisa, Italy}

\maketitle\abstracts{
Recent improved measurements of $B$ decays probing the $b\to s$ flavour-changing neutral current have put strong constraints on flavour violation beyond the Standard Model.
This talk reviews a model-independent analysis of these decays, which allows to put constraints on dimension-six $\Delta F=1$ effective operators.
These constraints can be used in turn to test the flavour structure of any theory beyond the SM.
}

\section{Introduction}

In the Standard Model (SM), flavour-changing neutral currents are GIM- and CKM-suppressed. As a consequence, they are sensitive to new physics. A particularly promising class of processes are radiative, semi-leptonic and leptonic $\Delta B=\Delta S=1$ decays, including the inclusive modes $B\to X_s\gamma$ and $B\to X_s\ell^+\ell^-$, the exclusive ones $B\to K^*\gamma$, $B\to K\mu^+\mu^-$ and $B\to K^*\mu^+\mu^-$ and the leptonic one $B_s\to\mu^+\mu^-$.
Contributions from physics beyond the SM to the observables in all these decays can be described by the modification of Wilson coefficients of local operators in an effective Hamiltonian of the form
\begin{equation}
\label{eq:Heff}
{\mathcal H}_{\text{eff}} = - \frac{4\,G_F}{\sqrt{2}} V_{tb}V_{ts}^* \frac{e^2}{16\pi^2}
\sum_i
(C_i O_i + C'_i O'_i) + \text{h.c.}
\end{equation}
In many concrete models, the operators that are most sensitive to new physics (NP) are a subset of the following ones,
\begin{align}
\label{eq:O7}
O_7^{(\prime)} &= \frac{m_b}{e}
(\bar{s} \sigma_{\mu \nu} P_{R(L)} b) F^{\mu \nu},
&
O_8^{(\prime)} &= \frac{g m_b}{e^2}
(\bar{s} \sigma_{\mu \nu} T^a P_{R(L)} b) G^{\mu \nu \, a},
\nonumber\\
O_9^{(\prime)} &= 
(\bar{s} \gamma_{\mu} P_{L(R)} b)(\bar{\ell} \gamma^\mu \ell)\,,
&
O_{10}^{(\prime)} &=
(\bar{s} \gamma_{\mu} P_{L(R)} b)( \bar{\ell} \gamma^\mu \gamma_5 \ell)\,,
\nonumber\\
O_S^{(\prime)} &= 
\frac{m_b}{m_{B_s}} (\bar{s} P_{R(L)} b)(  \bar{\ell} \ell)\,,
&
O_P^{(\prime)} &=
\frac{m_b}{m_{B_s}} (\bar{s} P_{R(L)} b)(  \bar{\ell} \gamma_5 \ell)\,,
\end{align}
denoted as (chromo-)magnetic, semi-leptonic and (pseudo-)scalar operators.
While the radiative $b\to s\gamma$ decays are sensitive only to the magnetic and chromomagnetic operators, semi-leptonic $b\to s\ell^+\ell^-$ decays are in principle sensitive to all the above operators\footnote{Since $C_7$ and $C_8$ contribute to all observables in a fixed linear combination, constraints on $C_8$ will not be discussed separately in the following.}. The scalar and pseudoscalar operators are most relevant for the $B_s\to\mu^+\mu^-$ decay.

\section{The impact of $B_s\to\mu^+\mu^-$}\label{sec:bsmm}

The decay $B_s\to\mu^+\mu^-$ is strongly helicity-suppressed in the SM. For this reason, its branching ratio could be strongly enhanced in the presence of NP in the scalar or pseudoscalar operators, which would lift this helicity suppression. A prominent example of a model predicting such enhancement is supersymmetry with large $\tan\beta$ and sizable $A$ terms, as motivated e.g.\ by grand unification.

\begin{figure}[tbp]
 \centering
 \includegraphics[width=10.5cm]{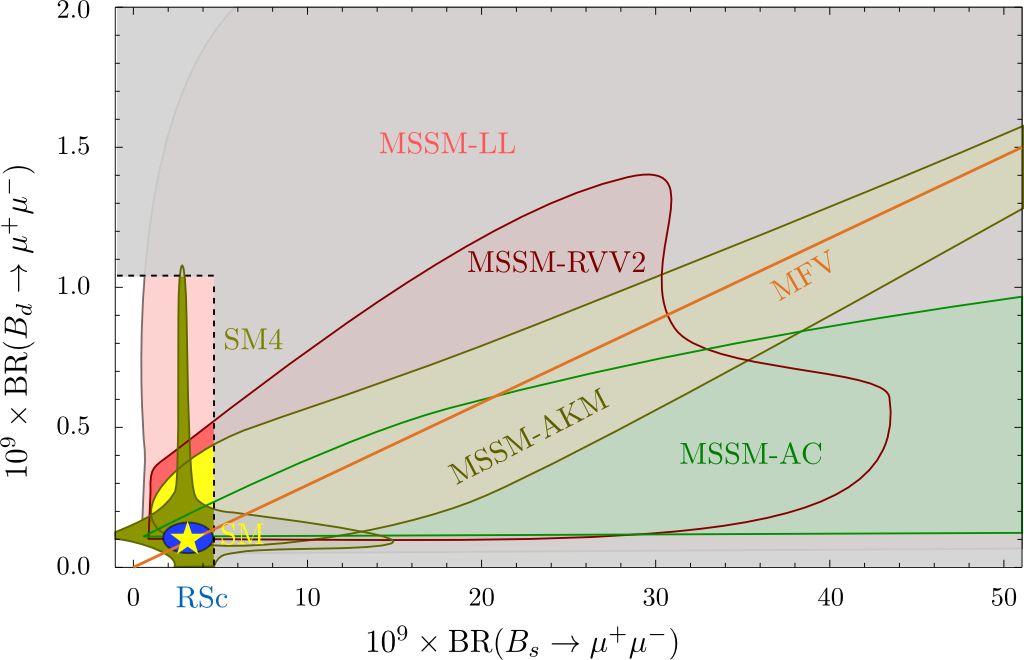}
 \caption{Correlation between the branching ratios of $B_s\to\mu^+\mu^-$ and $B_d\to\mu^+\mu^-$ in MFV, the SM4 and four SUSY flavour models. The gray area is ruled out experimentally. The SM point is marked by a star.}
 \label{fig:bqmumu}
\end{figure}

However, the recent upper bound on the branching ratio presented by the CMS collaboration\cite{Chatrchyan:2012rg} and the very recent, even stronger bound by LHCb presented at this conference\cite{Aaij:2012ac}, strongly limit the size of such contributions. This constitutes a significant constraint for a large class of NP models, as is exemplified in fig.~\ref{fig:bqmumu}, showing 
the correlation between $\text{BR}(B_s\to\mu^+\mu^-)$ and $\text{BR}(B_d\to\mu^+\mu^-)$ in models with Minimal Flavour Violation (MFV \cite{D'Ambrosio:2002ex}),
the Randall-Sundrum model with custodial protection (RSc \cite{Blanke:2008yr}), the Standard Model with a sequential fourth generation (SM4 \cite{Buras:2010pi}) and four SUSY flavour models\footnote{%
The acronyms stand for the models by Agashe and Carone (AC \cite{Agashe:2003rj}), Ross, Velasco-Sevilla and Vives (RVV2 \cite{Ross:2004qn}), Antusch, King and Malinsky (AKM \cite{Antusch:2007re}) and a model with left-handed currents only (LL \cite{Hall:1995es}). See the original analysis\cite{Altmannshofer:2009ne} for details.}
A large part of the parameter space of the supersymmetric models, where $\tan\beta$ can be large, is ruled out by the constraints, leading to a much more constrained situation than one year ago\cite{Straub:2010ih,Straub:2011gt}.
However, it should be emphasized that models where NP enters $B_s\to\mu^+\mu^-$ via the semi-leptonic operators $O_{10}^{(\prime)}$, like the SM4 or RSc in fig.~\ref{fig:bqmumu}, or SUSY models with small $\tan\beta$, are starting to be probed only now. Indeed, a model-independent analysis of new physics in $b\to s$ transitions has shown that NP in $C_{10}$ or $C_{10}'$ can only enhance the branching ratio of $B_s\to\mu^+\mu^-$ up to $5.6\times 10^{-9}$, using all the information on $b\to s$ transitions available before this conference\cite{Altmannshofer:2011gn}.

In any case, an important consequence of the strong new bounds is that
the scalar and pseudoscalar operators are irrelevant\footnote{Barring a fortuitous cancellation in $C_S-C_S'$ and $C_P-C_P'$, which are the only combinations entering the $B_s\to\mu^+\mu^-$ branching ratio.} for all the semi-leptonic $b\to s$ decays, which are not helicity suppressed. The following model-independent discussion will thus focus on the magnetic and semi-leptonic operators.

\section{Constraints on Wilson coefficients}

The wealth of experimental information on $b\to s$ transitions can be used to put constraints on physics beyond the SM.
Since the set of relevant operators and the dependence on the Wilson coefficients can be different for the various experimental observables probing the Hamiltonian (\ref{eq:Heff}), a combined analysis of all available experimental constraints is mandatory to obtain meaningful bounds on the individual Wilson coefficients and to determine the room left for NP. Such analyses have been performed e.g. for the magnetic penguin operators\cite{DescotesGenon:2011yn} or the SM operator basis\cite{Bobeth:2011gi,Bobeth:2011nj}. Recently, a comprehensive analysis of constraints on the Wilson coefficients $C_{7,9,10}^{(')}$, considering both inclusive and exclusive $b\to s$ transitions, has been published\cite{Altmannshofer:2011gn}.

\begin{figure}[tb]
\includegraphics[width=0.33\textwidth]{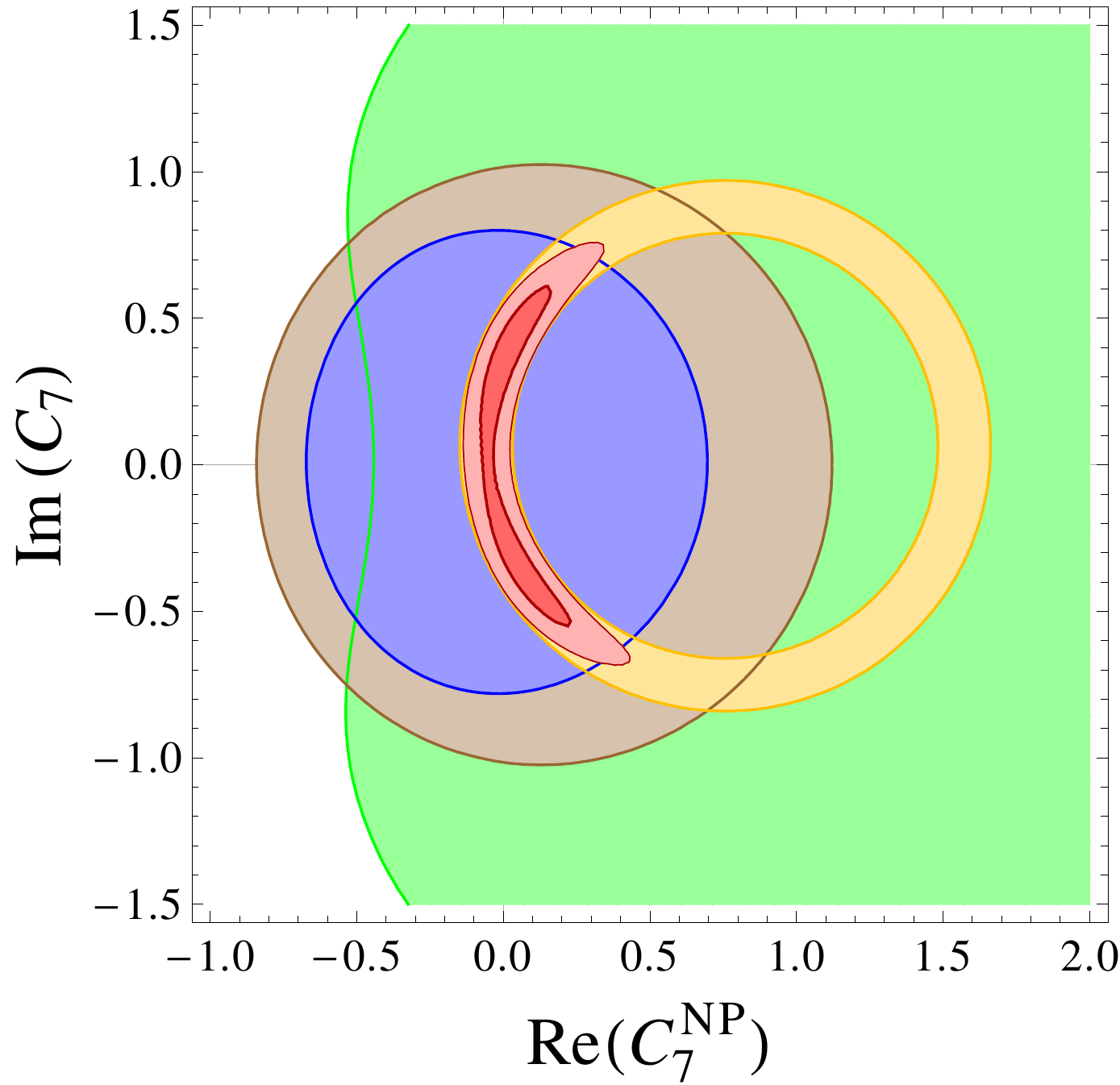}
\includegraphics[width=0.33\textwidth]{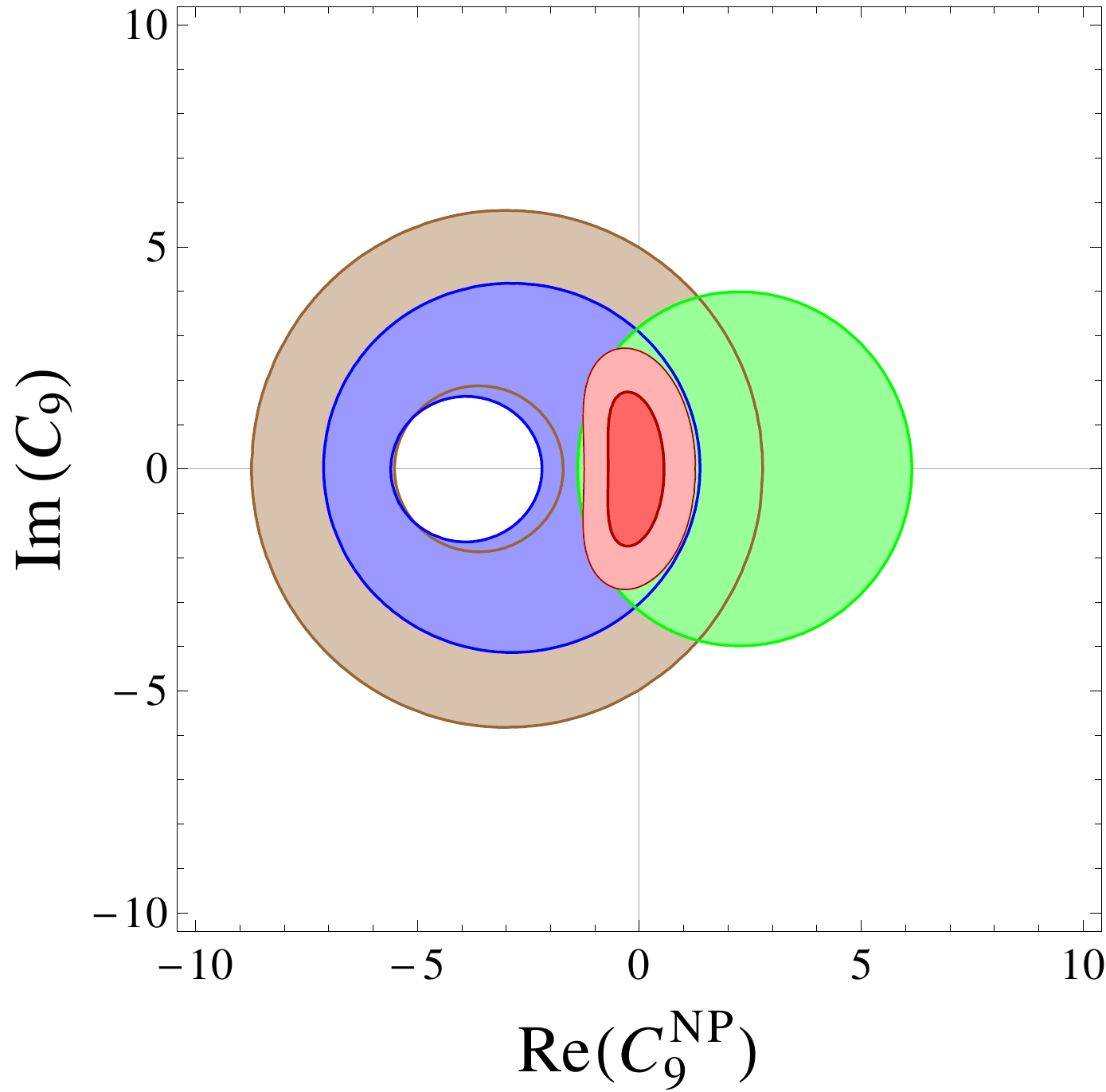}
\includegraphics[width=0.33\textwidth]{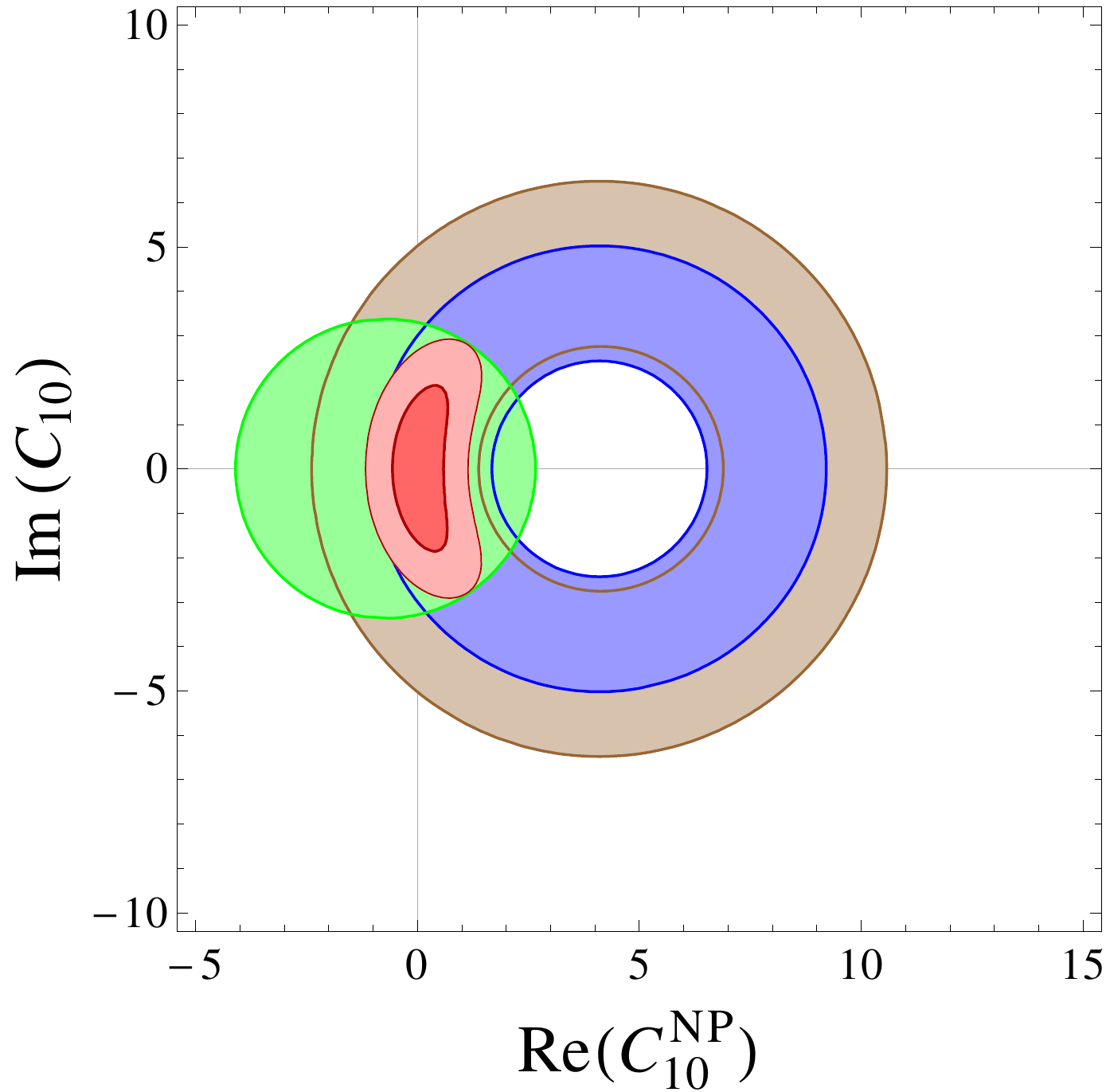}
\caption{Individual $2\sigma$ constraints in the complex planes of Wilson coefficients, coming from
$B\to X_s\ell^+\ell^-$ (brown), $B\to X_s\gamma$ (yellow), $A_\text{FB}(B\to K^*\mu^+\mu^-)$ (green) and BR$(B\to K^*\mu^+\mu^-)$ (blue), as well as combined 1 and $2\sigma$ constraints (red).}
\label{fig:wcconstraints}
\end{figure}

\begin{figure}[tb]
\includegraphics[width=0.33\textwidth]{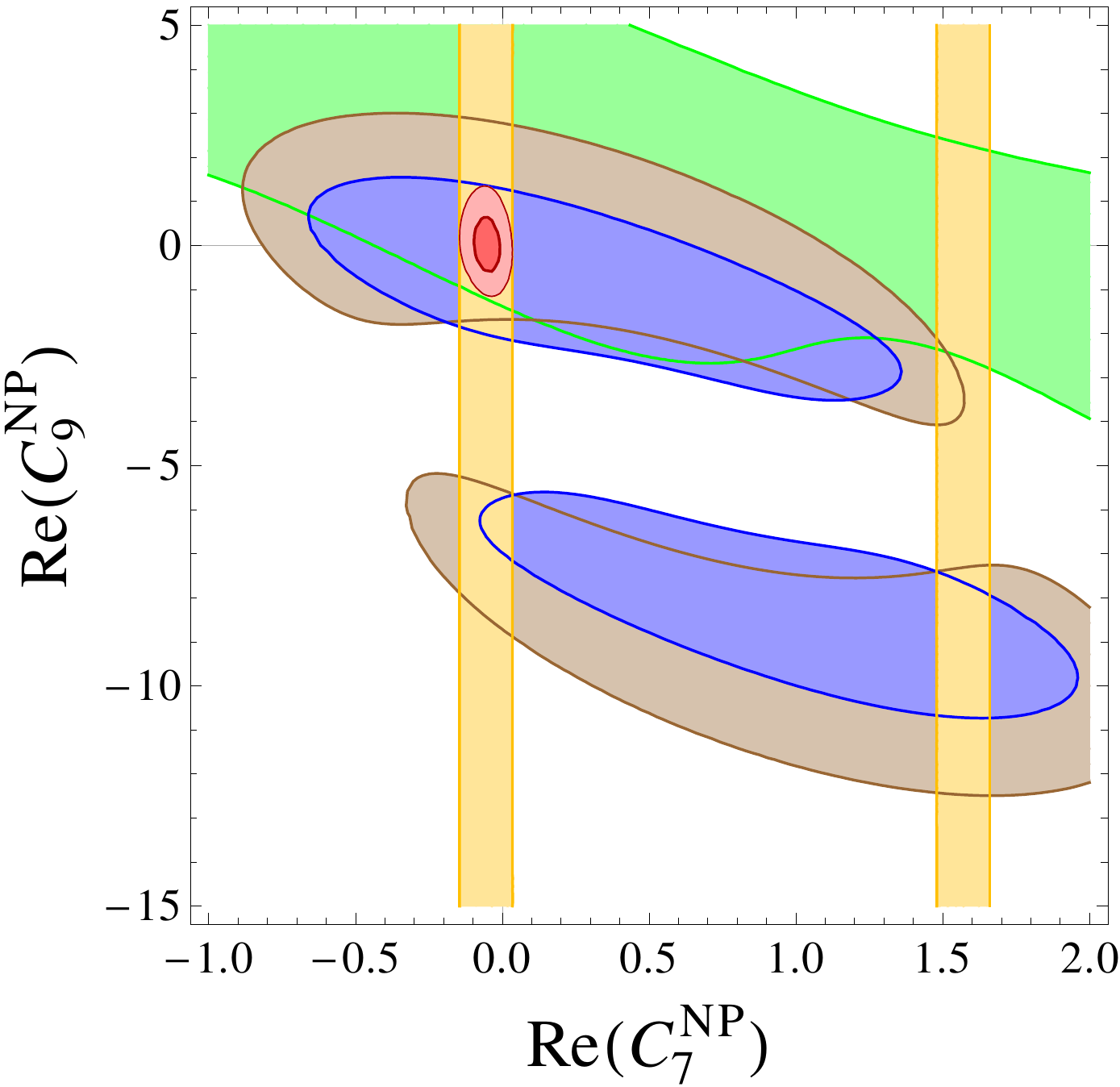}
\includegraphics[width=0.33\textwidth]{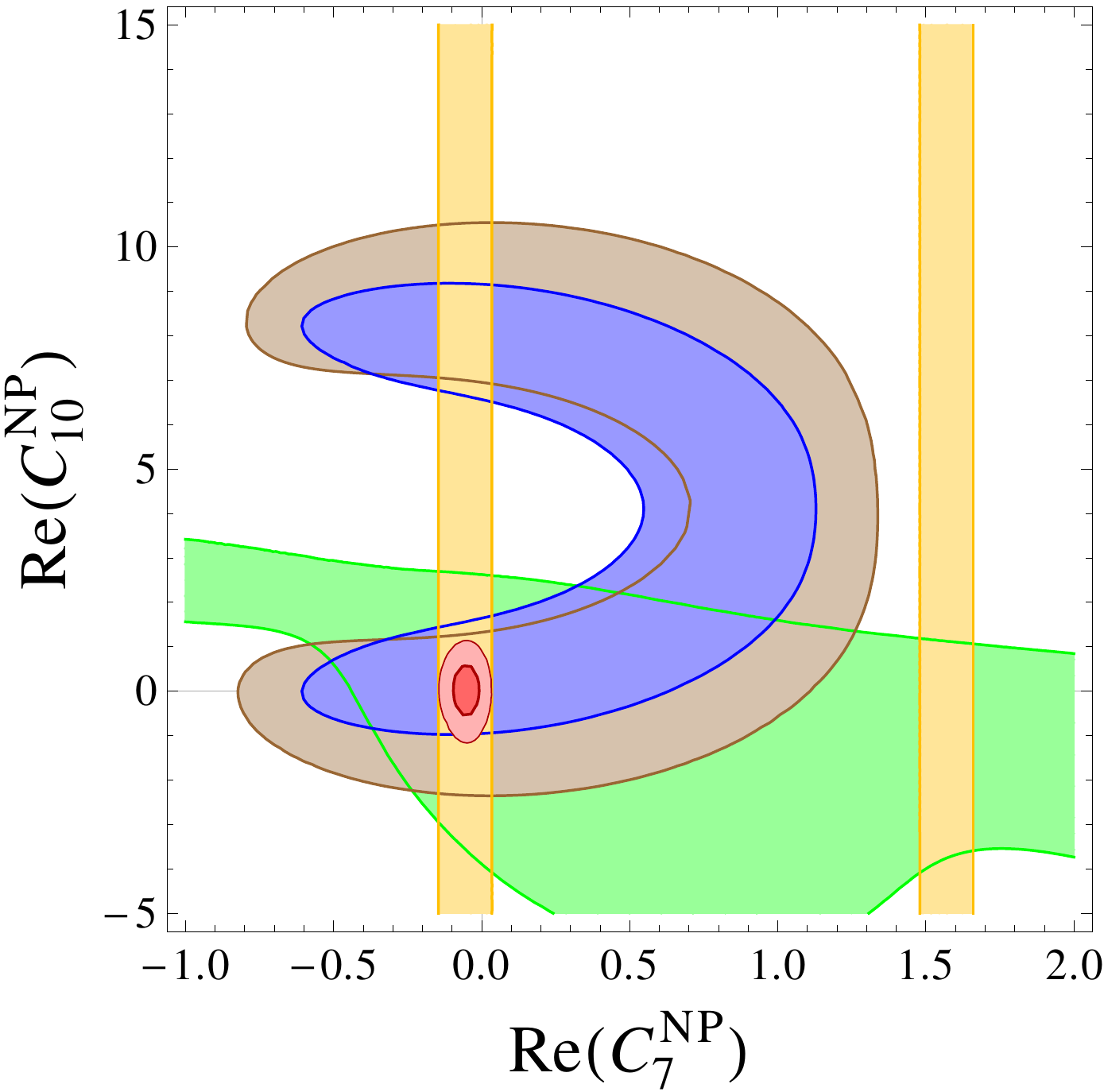}
\includegraphics[width=0.33\textwidth]{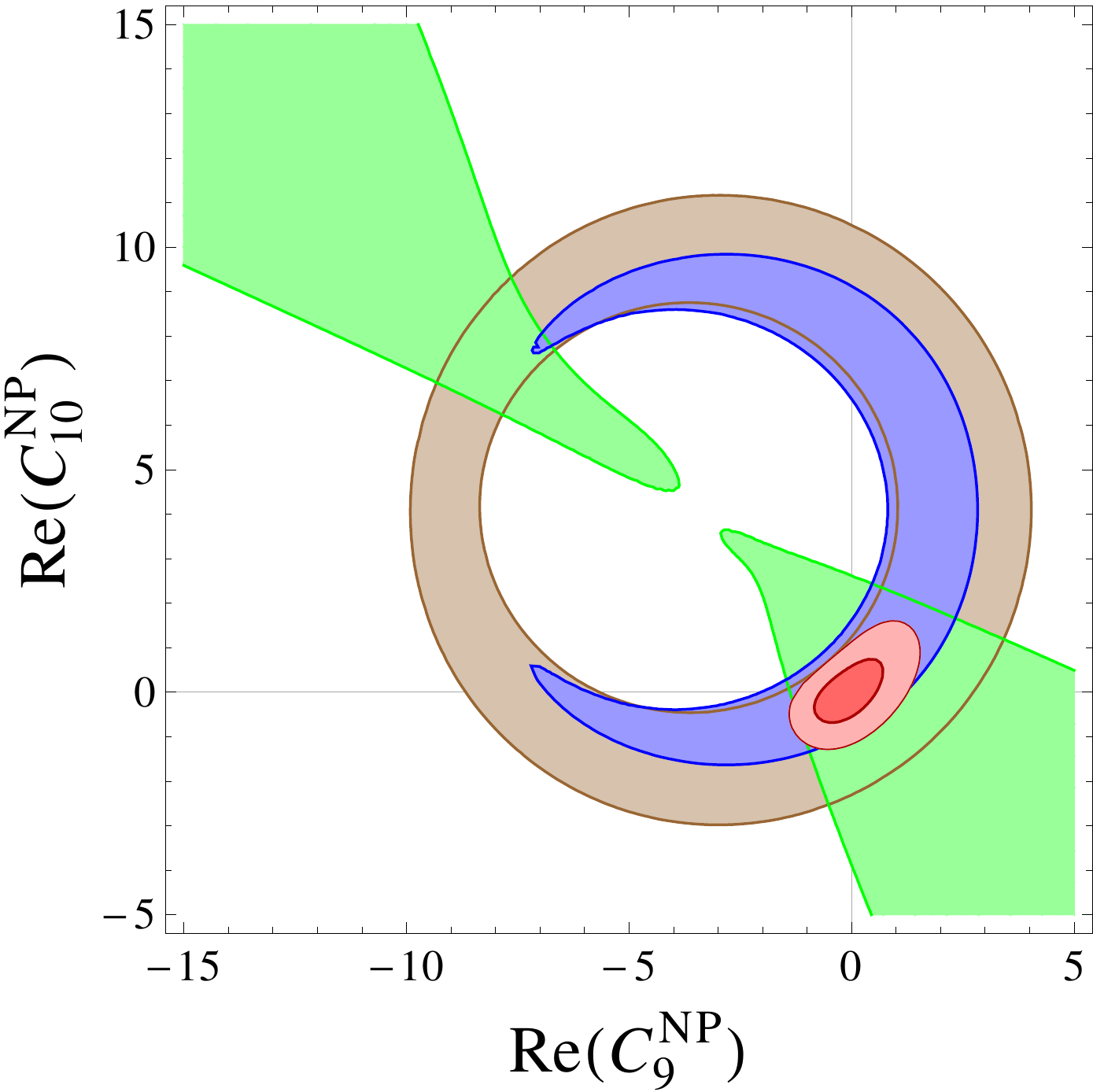}
\caption{Individual $2\sigma$ constraints on the correlation between real Wilson coefficients. Colour coding same as in fig.~2.
}
\label{fig:wcconstraints2}
\end{figure}

Based on this study, figs.~\ref{fig:wcconstraints} and \ref{fig:wcconstraints2} show the constraints on the Wilson coefficients $C_{7,9,10}$ from individual processes as well as the combined constraints. For these plots, only two coefficients at a time (or the real and imaginary parts of one coefficient) were varied, while the other coefficients were fixed to their SM values. The observables considered include
\begin{itemize}
\item the branching ratio of the inclusive $B \to X_s \gamma$ decay (the 95\% C.L. constraint from this observable alone is shown in figs.~\ref{fig:wcconstraints} and \ref{fig:wcconstraints2} as a yellow band),
\item the partial branching ratio of the inclusive $B \to X_s \ell^+ \ell^-$ decay both at low and high dilepton invariant mass $q^2$ (in brown),
\item the mixing induced CP asymmetry in $B\to K^*\gamma$, 
which is sensitive to the chirality-flipped Wilson coefficient $C_7'$ (not relevant for figs.~\ref{fig:wcconstraints} and \ref{fig:wcconstraints2}),
\item the partial branching ratio (blue), forward-backward asymmetry $A_\text{FB}$ (green) and $K^*$ longitudinal polarization fraction $F_L$ (not shown individually, but included in the combination) in $B \to K^* \mu^+ \mu^-$, both at low and high $q^2$. 
\end{itemize}
One can make the following observations.
\begin{itemize}
\item At the 95\% C.L., all best fit regions are compatible with the SM.
\item The combination of inclusive and exclusive $b\to s\ell^+\ell^-$
      observables exclude sign flips in various low-energy Wilson coefficients.
      That is, the SM is likely to provide the dominant effects
      in low energy observables. To arrive at this conclusion, high-$q^2$ data on $B\to K^*\ell^+\ell^-$ are competitive with and coplementary to the low-$q^2$ ones.
\item The constraints on the imaginary parts of $C_7$, $C_9$ and $C_{10}$ are looser than on the real parts.
This can be understood from the fact that in the branching ratios and CP averaged angular observables giving the strongest constraints, only NP contributions aligned in phase with the SM can interfere with the SM contributions. As a consequence, NP with non-standard CP violation is in fact constrained more weakly than NP where CP violation stems only from the CKM phase. This highlights the need for improved measurements of CP asymmetries directly sensitive to non-standard phases.
\end{itemize}

Similar constraints can be obtained\cite{Altmannshofer:2011gn} for the chirality-flipped Wilson coefficients $C_{7,9,10}'$. Finally, a global analysis, varying all the coefficients simultaneously, shows that sizeable contributions to individual coefficients cannot be excluded yet in general\cite{Altmannshofer:2011gn}, due to cancellations among different coefficients.

\section{Outlook}

Using the global analysis of Wilson coefficients, allowed ranges for observables to be measured in the future can be obtained. By means of the generality of this approach, these conclusions are valid {\em for any theory} beyond the SM described by the Hamiltonian~(\ref{eq:Heff}) and are useful to assess the prospects of future measurements. Table~\ref{tab:pred} summarizes such predictions for the branching ratio of $B_s\to\mu^+\mu^-$ -- assuming no NP in (pseudo-)scalar operators -- and for various low-$q^2$ angular observables in $B\to K^*\mu^+\mu^-$.
The allowed range for BR($B_s\to\mu^+\mu^-$) shows that the new, tight constraints are just starting to become sensitive to NP in semi-leptonic operators, as discussed in sec.~\ref{sec:bsmm}.
In the presence of non-standard CP violation, the low-$q^2$ angular CP asymmetries $A_7$ and $A_8$ in $B\to K^*\mu^+\mu^-$ can reach up to $\pm35\%$ and $\pm20\%$, respectively.
In the presence of right-handed currents, the angular observables $A_9$ and $S_3$ in $B\to K^*\mu^+\mu^-$ can reach up to $\pm15\%$ at low $q^2$.

The prospects for improved sensitivity to NP in $b\to s$ transitions in the near future are excellent. While the $B$ factories still have potential to improve the analyses of inclusive decays, LHCb is expected to strongly improve its precision of $B\to K^{(*)}\mu^+\mu^-$ observables. This has been impressively demonstrated with the new results presented at this conference\cite{LHCb-CONF-2012-008}.
In the near future, particularly interesting observables in exclusive decays will be the angular CP asymmetries $A_7$, $A_8$ and $A_9$ as well as the CP-averaged $S_3$, $S_4$ and $S_5$ in $B\to K^*\mu^+\mu^-$, the branching ratio and angular observables in $B\to K\mu^+\mu^-$, the mixing-induced CP asymmetry in $B\to K^*\gamma$, and the branching ratios of $B_{s,d}\to\mu^+\mu^-$. These measurements have the potential to uncover the first signs of new physics in $b\to s$ transitions or, if no deviation from the SM expectations is found, to put even stronger constraints on physics beyond the SM.

\begin{table}[t]
\addtolength{\arraycolsep}{10pt}
\renewcommand{\arraystretch}{1.6}
\begin{center}
\begin{tabular}{|l|ccccc|}
\hline
Scenario & BR($B_s\to\mu^+\mu^-$) & $|\langle A_7\rangle_{[1,6]}|$ & $|\langle A_8\rangle_{[1,6]}|$ & $|\langle A_9\rangle_{[1,6]}|$ & $\langle S_3\rangle_{[1,6]}$  \\
\hline
Real LH & $[1.0,5.6] \times 10^{-9}$ & 0 & 0 & 0 & 0 \\
Complex LH & $[1.0,5.4] \times 10^{-9}$ & $<0.31$ & $<0.15$ & 0 & 0 \\
Complex RH & $<5.6\times 10^{-9}$  & $<0.22$ & $<0.17$ & $<0.12$ & $[-0.06,0.15]$ \\
Generic NP & $<5.5\times 10^{-9}$  & $<0.34$ & $<0.20$ & $<0.15$ & $[-0.11,0.18]$\\
\hline
\end{tabular}
\end{center}
\caption{Predictions at 95\% C.L. for the branching ratio of $B_s\to\mu^+\mu^-$ and for low-$q^2$ angular observables in $B\to K^*\mu^+\mu^-$ (neglecting tiny SM effects
below the percent level) in four scenarios with real or complex NP effects in $C_i$ only (LH), $C_i'$ only (RH) or both (generic NP), assuming negligible (pseudo-)scalar currents.}
\label{tab:pred} 
\end{table} \normalsize

\section*{Acknowledgements}
I thank the organizers of the Moriond EW Session for the opportunity to contribute to this inspiring conference. I also thank my collaborators Wolfgang Altmannshofer and Paride Paradisi for the pleasant and fruitful collaboration and the former for proofreading the manuscript. This work was supported by the EU ITN ``Unification in the
LHC Era'', contract PITN-GA-2009-237920 (UNILHC).

\section*{References}
\raggedright

\end{document}